\documentclass[%
 reprint,
 amsmath,amssymb,
 aps,
]{revtex4-2}
\usepackage[dvipsnames]{xcolor}
\usepackage{soul}
\usepackage{graphicx}
\usepackage{dcolumn}
\usepackage{bm}
\usepackage[mathcal]{eucal}

\begin{document}

\preprint{APS/123-QED}

\title{Noise factor of Brillouin amplifiers}
\author{John H. Dallyn$^{1,2,3}$}
\email{jhdally@sandia.gov}
\author{Nils T. Otterstrom$^1$}
\author{Matt Eichenfield$^{1,4}$}
\author{Peter T. Rakich$^5$}
\author{Ryan O. Behunin$^{2,3}$}
\affiliation{%
 $^1$Microsystems Engineering, Science, and Applications,
Sandia National Laboratories, Albuquerque, New Mexico 87123, USA 
}%
\affiliation{%
 $^2$Department of Applied Physics and Material Sciences, Northern Arizona University, Flagstaff, Arizona 86011, USA
}%
\affiliation{%
$^3$Center for Materials Interfaces in Research and Applications (¡MIRA!), Flagstaff, Arizona 86011, USA
}%
\affiliation{%
 $^1$Microsystems Engineering, Science, and Applications,
Sandia National Laboratories, Albuquerque, New Mexico 87123, USA 
}%
\affiliation{%
 $^4$College of Optical Sciences, University of Arizona, Tucson, Arizona 85719, USA 
}%
\affiliation{%
 $^5$Department of Applied Physics, Yale University, New Haven, Connecticut 06520, USA 
}%

\date{\today}

\begin{abstract}
Stimulated Brillouin scattering (SBS), an optical nonlinearity arising from photon-phonon interactions, has formed the basis for a large class of optical signal processing devices, including Brillouin amplifiers. A limiting factor of such amplifiers is the noise due to thermal-mechanical fluctuations that the phonons imprint on the optical signal. Prior work has either inferred or experimentally observed a noise factor ($F$) that depends only on the thermal occupation of the phonons ($F\approx	1+n_{th}$). We show that this noise factor results naturally from a Hamiltonian-based spatio-temporal coupled mode treatment in the limit of large Brillouin amplification and when phonon propagation is neglected. Moreover, this theoretical framework allows us to extend our treatment to a much larger and more representative parameter space for emerging SBS systems; specifically, this analysis accounts for the forward or backward nature of the scattering process and the effects of phonon propagation, optical loss, and small Brillouin gains. Our results demonstrate that the noise factor can deviate radically from $F\approx	1+n_{th}$ for a host of modern SBS devices, especially those in which phonon propagation significantly changes the coupled mode dynamics.
\end{abstract}

\maketitle

\section{Introduction}

\indent Stimulated Brillouin scattering (SBS), arising from photon-phonon interactions in traveling-wave systems, has emerged as a powerful mechanism for photonic-based optical amplification, with large, exponential gain \citep{boyd2008nonlinear} and highly tunable system parameters\citep{rakich2012giant,shin2013tailorable}. Given these properties, SBS has enabled a set of advanced optical signal processing capabilities in fiber\citep{merklein2022100} and chip-scale systems\citep{eggleton2019brillouin}, including high-gain optical amplifiers\citep{ippen1972stimulated,olsson1987characteristics,pant2011chip,rakich2012giant,kittlaus2016large,kittlaus2017chip}, isolators\citep{huang2011complete,poulton2012design,kim2015non,kittlaus2018non}, ultra-low noise lasers\citep{smith1991narrow,geng2006highly,grudinin2009brillouin,li2012characterization,li2013microwave,kabakova2013narrow,li2014low,loh2015noise,morrison2017compact,behunin2018fundamental,otterstrom2018silicon,gundavarapu2019sub,dallyn2022thermal}, and tunable optical filters\citep{tanemura2002narrowband,vidal2006tunable,zadok2007gigahertz,braje2009brillouin,wise2011sharp,zhang2012ultrawide,dong2015brillouin,marpaung2015low}. However, because these processes rely on GHz-frequency phonon modes, thermal-mechanical fluctuations---inevitably present in SBS processes at non-zero temperatures---are imprinted on the optical signals of interest, resulting in optical noise that profoundly impacts the performance and use cases of these SBS technologies. 
\newline   
\indent  Noise in amplifier systems is quantified by the noise factor, which captures how the signal-to-noise ratio degrades in the amplification process and can be quantified by comparing the additive amplifier noise with the amplified input noise. RF amplifier systems are benchmarked by thermal input noise, as the scale of the thermal energy is significantly larger than the quantum energy ($k_B T\gg \hbar \omega$). By contrast, in the case of optical amplifiers, $\hbar \omega \gg k_B T$, meaning that the intrinsic quantum EM field fluctuations dominate\cite{haus1998noise}. As such, the appropriate input noise upon which to calibrate the noise factor is zero-point noise due to vacuum fluctuations\cite{haus1998noise}. While the noise figure of traditional Brillouin systems has been inferred through measurement for a relatively narrow set of conditions\citep{olsson1987characteristics,desurvire2002erbium}, to our knowledge, this ultimate limit of the SBS noise factor has yet to receive a general theoretical treatment. Moreover, the rapid advances in the field of Brillouin photonics has unveiled numerous systems whose dynamics are not adequately captured by traditional approximations, highlighting the pressing need for a comprehensive and accessible noise figure formulation.
\newline
\indent In this paper, we present a traveling wave-based Hamiltonian model \citep{sipe2016hamiltonian,kharel2016noise} that captures the relevant noise, coupling, and field evolution in a Brillouin amplifier under a broad range of conditions. For a large class of Brillouin systems, the phonon decay rate is large enough to neglect propagation of the phonons, what we call the non-propagating phonon (NPP) approximation. This is reflected in previous theoretical models that assume the phonon group velocity, $v_{g,b}$, is equal to zero\citep{boyd2008nonlinear}. We show that in this regime, along with large Brillouin amplification, the resulting noise factor is $F\approx 1+n_{th}$, consistent with experimental and inferred noise factor\citep{olsson1987characteristics,desurvire2002erbium,otterstrom2020shaping}. However, the NPP approximation overlooks a range of emerging systems\citep{kharel2019high,kharel2022multimode,yoon2023simultaneous,otterstrom2023modulation} where phonon propagation is substantial. In this regime, we find that accounting for phonon propagation results in a markedly reduced noise figure compared to conventional understanding.
\newline
\indent The paper is organized as follows: Section \ref{sec:theory} describes the theory of the noise factor for a linear Brillouin optical amplifier when the NPP approximation is valid. Section \ref{sec:Full} presents the noise factor for forward intermodal and backward Brillouin scattering, assuming  $v_{g,b}$ is nonzero, and compares with results obtained using the NPP approximation. Section \ref{sec:discuss} summarizes and discusses the key results of the paper. 

\section{Theory}\label{sec:theory}
Large SBS-based optical amplification occurs when a Stokes signal wave with frequency $\omega_S$ and wave vector $\mathbf{k}_{S}$ is red-detuned from a strong pump wave ($\omega_p$, $\mathbf{k}_p$) by the mechanical Brillouin frequency ($\Omega_m$, $\mathbf{q}_m$), which is determined by phase matching requirements, i.e., energy conservation ($\omega_p=\omega_S+\Omega_m$) and momentum conservation ($\mathbf{k}_p=\mathbf{k}_S+\mathbf{q}_m$). Under these conditions, the pump and Stokes waves produce periodic optical forces that are resonant with a traveling-wave phonon mode, resulting in dynamical Bragg scattering, which transfers energy from the pump wave to the red-detuned Stokes wave.   
\newline
\indent To calculate the noise factor of a Brillouin amplifier, we utilize the Hamiltonian, $H = H_0+H_{int}$ for a Stokes Brillouin interaction between two continuous wave optical fields, where $H_0$ governs the uncoupled pump ($A_p$), Stokes ($A_s$), and phonon ($B$) envelopes \cite{sipe2016hamiltonian,kharel2016noise},
\begin{equation}\label{Ham0}
\begin{split}
H_0= \hbar \int dz \bigl[&A^{\dagger}_p(z,t) \hat{\omega}_p A_p(z,t)+A^{\dagger}_s(z,t) \hat{\omega}_s A_s(z,t)\\
&+B^{\dagger}(z,t) \hat{\Omega}_z B(z,t)\bigr],
\end{split}
\end{equation}
and $H_{int}$ describes the Brillouin interaction between these envelopes \cite{sipe2016hamiltonian,kharel2016noise},
\begin{equation}\label{HamInt}
H_{int}= \hbar \int dz \bigl[g_0 A^{\dagger}_p(z,t) A_s(z,t) B(z,t)\bigr] e^{i(q_m-\Delta k_s)z}+H.c.
\end{equation}
Here $g_0$ is the distributed optomechanical coupling, $q_m$ is the phonon wave vector, and $\Delta k_s$ is the spatial frequency of the optical beat note ($\Delta k_s= k_p - k_s$). If the three fields are co-propagating, then $k_s>0$. If the Stokes field is counterpropagating to the pump and phonon fields, then $k_s<0$. The spatial operators of the phonon, pump, and Stokes fields are $\hat{\Omega}_z \simeq \Omega_m-i v_{g,b} \partial_z$, $\hat{\omega}_p \simeq \omega_p-i v_{g,p} \partial_z$, and $\hat{\omega}_s \simeq \omega_s-i v_{g,s} \partial_z$, where the slowly varying envelope approximation permits higher order of dispersion to be neglected \cite{kharel2016noise}. The relative signs of $v_{g,b}$, $v_{g,p}$, and $v_{g,s}$ determine whether the interaction is forward intermodal or backward Brillouin scattering. 
\newline
\indent When spatial phase matching is satisfied, which is the case we will consider going forward, then $q_m-\Delta k_s=0$. The resultant Heisenberg-Langevin equations of motion are given by \cite{sipe2016hamiltonian,kharel2016noise}

\begin{align} 
 \frac{\partial \overline{B}}{\partial t} &= -i \bigl(\Omega_m - \Omega\bigr)
  \overline{B}-\frac{\Gamma}{2} \overline{B} - v_{g,b} \frac{\partial  
  \overline{B}}{\partial z} - i g_0^* \overline{A}_s^\dagger \overline{A}_p +  
  \eta \label{Beom} 
  \\ 
 \frac{\partial \overline{A}_p}{\partial t} &= -\frac{\gamma_p}{2}
  \overline{A}_p - v_{g,p} \frac{\partial  
  \overline{A}_p}{\partial z} - i g_0 \overline{A}_s \overline{B} + \xi_p 
  \label{Apeom}  
  \\
 \frac{\partial \overline{A}_s}{\partial t} &= -\frac{\gamma_s}{2}
  \overline{A}_s - v_{g,s} \frac{\partial  
  \overline{A}_s}{\partial z} - i g_0^* \overline{A}_p \overline{B}^\dagger +  
  \xi_s. \label{Aseom} 
\end{align}
Here, we have moved our fields into the rotating frame with the slowly varying envelopes $\overline{B}(z,t) = B(z,t) e^{i \Omega t}$, $\overline{A}_p(z,t) = A_p(z,t) e^{i \omega_p t}$, and $\overline{A}_s(z,t) = A_s(z,t) e^{i \omega_s t}$, where $\Omega = \omega_p - \omega_s$. We have included decay rates of $\Gamma$, $\gamma_p$, and $\gamma_s$, with corresponding Langevin forces of $\eta$, $\xi_p$, and $\xi_s$ for the phonon, pump, and Stokes fields, respectively. We assume that the decay rates and the Langevin forces act together to drive the system back to thermal equilibrium when external driving is no longer applied, according to the fluctuation-dissipation theorem. The Langevin forces are assumed to be zero-mean Gaussian variables
with white power spectra, whose two-time correlation properties are defined as\citep{kharel2016noise}
\begin{align}
\langle&\eta^{\dagger}(z_1,t) \eta(z_2,t^{\prime})\rangle=\Gamma n_{th} \delta\left(t-t^{\prime}\right) \delta(z_1-z_2),\label{phononlang}\\
\langle&\xi_s^{\dagger}(z_1,t) \xi_s(z_2,t^{\prime})\rangle=\gamma_s N_{th} \delta\left(t-t^{\prime}\right) \delta(z_1-z_2)\label{photonlangHC}\\ 
\langle&\eta(z_1,t) \eta^{\dagger}(z_2,t^{\prime})\rangle=\Gamma (n_{th}+1) \delta\left(t-t^{\prime}\right) \delta(z_1-z_2),\label{phononlangHC}\\
\langle&\xi_s(z_1,t) \xi_s^{\dagger}(z_2,t^{\prime})\rangle=\gamma_s(N_{th}+1)  \delta\left(t-t^{\prime}\right) \delta(z_1-z_2)\label{photonlang}
\end{align}
where ${N_{th}=({\rm exp}(\hbar\omega_s/k_B T_0)-1)^{-1}}$ and ${n_{th}=({\rm exp}(\hbar\Omega/k_B T_0)-1)^{-1}}$ are the thermal occupation numbers of the optical and acoustic modes, respectively, $T_0$ is the temperature, and $\hbar$ and $k_B$ are the Planck and Boltzmann constants. Cross-correlations between different Langevin forces are assumed to be zero. Going forward, we will neglect the contribution of the optical thermal occupation number as it is vanishingly small at relevant temperatures. Additionally, we assume the pump is undepleted (i.e., $\overline{A}_p(z,t)$ is constant in space and time) and there is no external pump noise. While pump noise can be accounted for in the undepleted pump approximation, we are focusing on the noise due to the thermal phonon field in this work, as it is the predominant source of noise in most traveling-wave optomechanical systems. In this limit the coupled mode equations become linear and can be solved using standard methods. We Fourier transform  Eqs. \ref{Beom} and \ref{Aseom} to express these equations in the frequency domain, giving
\begin{align} 
 &\biggl[i\bigl(\Omega - \Omega_m -\omega\bigr) + \frac{\Gamma}{2}\biggr]\overline{B}
  (z,\omega)+v_{g,b}\frac{\partial}{\partial z}\overline{B}
  (z,\omega)\nonumber\\ & =  - i g_0^* 
  \overline{A}_p \overline{A}_s^\dagger(z,\omega) + \eta(z,
  \omega)\label{BTreomfullmain} \\ 
 &\Bigl(-i\omega + \frac{\alpha v_{g,s}}{2}\Bigr)\overline{A}_s(z,\omega)+v_{g,s}\frac{\partial}{\partial z}\overline{A}_s(z,\omega)
  \nonumber\\ &  =  - i g_0^* \overline{A}_p 
  \overline{B}^\dagger(z,\omega) + \xi_s(z,\omega). \label{AsTreomfullmain}
\end{align}
We have made the substitution $\gamma_s=\alpha v_{g,s}$, where $\alpha$ is the linear optical loss of the amplifier.
\newline
\indent In the following subsections, we will derive an equation for the Stokes' field amplitude ($\overline{A}_s$) and calculate the two-sided power spectral density (PSD) of the amplitude, $S_{A_s}(z,\omega)$, defined by
\begin{equation}\label{PSD}
S_{A_s}(z,\omega)= \int_{-\infty}^{\infty} d\tau \ e^{i \omega \tau}\bigl\langle \overline{A}_s(z,t+\tau) \overline{A}^\dagger_s(z,t)\bigr\rangle.
\end{equation}
The PSD contains terms describing the amplified input signal, thermal mechanical noise, and optical vacuum fluctuations, all of which are necessary to calculate the noise factor.
\newline
\indent For a large swath of Brillouin amplifier systems, the decay rate of the phonons is large enough that the spatial propagation of the phonons can be neglected, and the NPP approximation can be applied. This NPP approximation is taken by setting the phonon group velocity $v_{g,b}$ to zero \cite{boyd2008nonlinear}. Applying the NPP approximation to Eq. \ref{BTreomfullmain}, solving for $\overline{B}$, and taking the Hermitian conjugate, the phonon field amplitude is
\begin{equation}\label{BdagC1sim}
\overline{B}^\dagger(z,\omega) = \chi^*_B\bigl[i g_0 A_p^\dagger \overline{A}_s(z,\omega) + \eta^\dagger(z,\omega)\bigr],
\end{equation}
where the phonon susceptibility, $\chi_B$, is given by $\chi_B=\bigl[i(\Omega - \Omega_m -\omega) + \Gamma/2\bigr]^{-1}$.  Incorporating the NPP solution for the phonon field, the decoupled Stokes equation becomes 
\begin{align}\label{AsC1}
\frac{\partial}{\partial z}&\overline{A}_s(z,\omega)+\frac{1}{v_{g,s}}\chi\overline{A}_s(z,\omega)\nonumber\\ &= - \frac{i g_0^* \overline{A}_p\chi^*_B}{v_{g,s}}\eta^\dagger(z,\omega)+\frac{1}{v_{g,s}}\xi(z,\omega)
\end{align}
where $\chi= (i\omega + \alpha v_{g,s}/2)-v_{g,s}\chi_B G_B P \Gamma/4$ is the effective susceptibility of the Stokes mode. We have made the substitution $|g_0|^2|\overline{A}_p|^2 = v_{g,s} G_B P \Gamma/4$, where $G_B$ is the Brillouin gain and $P$ is the power of the pump\citep{kharel2016noise}. This substitution is made for greater connection of the theory to accessible experimental parameters. Equation \ref{AsC1} has an integral solution of 
\begin{align}\label{AsC1zw}
 &\overline{A}_s(z,\omega) = e^{-\chi^* z/v_{g,s}
  }\overline{A}_s(0,\omega) + \frac{1} 
  {v_{g,s}}\nonumber\\ &\times \int_0^z dz_1 \biggl[-i g_0^* \chi^*_B\overline{A}_p\eta^\dagger(z_1,
  \omega) + \xi_s(z_1,\omega)\biggr]\nonumber\\ &\qquad\qquad\times e^{-
  \chi^*(z-z_1)/v_{g,s}}.
\end{align}
The PSD of Eq. \ref{AsC1zw}, according to Eq. \ref{PSD} with Langevin correlations given by Eqs. \ref{phononlang} and \ref{photonlang}, is
\begin{align}\label{AssimPSD}
 &S_{A_s}(z,\omega) = e^{-2 \Re[\chi^*] z/v_{g,s}
  } \Bigl[S_{A_s}^{C}(0,\omega) + S_{A_s}^{N}(0,\omega)\Bigr]
  \nonumber\\ & + \frac{\alpha +  G_B P |
  \chi_B|^{2}\Gamma^2 n_{th}/4}{2 \Re\bigl[\chi^*\bigr]}\Bigl(1-
  e^{-2\Re[\chi^*] z/v_{g,s}}\Bigr)
\end{align}
where $S_{A_s}^{C}(0,\omega) + S_{A_s}^{N}(0,\omega)$ is the injected envelope decomposed into the coherent ($C$) and noise signals ($N$). We do this to capture the vacuum input noise and initial signal of the Stokes amplitude at $z=0$. This expression captures the amplification of the injected signal and noise, term 1, and the additional noise inherent to the Brillouin process in term 2. We have neglected noise contributions from optical thermal occupation ($N_{th}$), since at relevant temperatures these are exceedingly small.

\subsection{Noise factor in the NPP approximation}

To characterize the noise in our Brillouin systems, we define the noise factor ($F$) as a ratio of the input signal to noise ratio ($SNR_1$) and the output signal to noise ratio ($SNR_2$):
\begin{equation}\label{SNRratio}
F \equiv \frac{SNR_1}{SNR_2}.
\end{equation}
Following the method in Refs. \onlinecite{otterstrom2020shaping} and \onlinecite{otterstrom2023modulation}, we assume that the coherent component of the injected field is $S_{A_s}^{C}(0,\omega)=|A_s^{in}|^2\delta(\omega)$, where $A_s^{in}$ is the amplitude of the input optical signal at $z=0$. Physically, we are assuming a signal comprised of a single frequency and the spectrum has been centered at $\omega=0$. For a Brillouin system, the input noise component is $S_{A_s}^{N}(0,\omega)= 1/v_{g,s}$, which is vacuum noise, as discussed in the introduction and Ref. \onlinecite{haus1998noise}. Therefore, $SNR_1$ for this system is
\begin{equation}\label{SNR1}
SNR_1 = \frac{\int_{-\Delta\omega}^{\Delta\omega}d\omega' |A_s^{in}|^2\delta(\omega)}{\int_{-\Delta\omega}^{\Delta\omega}d\omega' \frac{1}{v_{g,s}}} = \frac{|A_s^{in}|^2v_{g,s}}{2\Delta\omega}.
\end{equation}
We have taken integrals over a bandwidth, $\Delta\omega$, that is much smaller that the Brillouin gain bandwidth, $\Gamma$, of the signal (numerator) and noise (denominator). We choose an in-band definition for the noise factor by setting $\omega=0$. We do this because the Brillouin gain bandwidth is exceptionally narrow compared to other amplifiers. Therefore, an out-of-band definition, where the noise is evaluated at a frequency different than the signal, can be misleading for Brillouin amplifiers, while still mathematically correct. The out-of-band thermal fluctuations will not experience gain and the resulting noise factor will be less than one. We use Eq. \ref{AssimPSD} to calculate the output signal to noise ratio ($SNR_2$). The term proportional to $S_{A_s}^{C}(0,\omega)$ represents the amplified signal and the remaining terms originate from noise, resulting in
\begin{align}
 &SNR_2 = \frac{|A_s^{in}|^2 v_{g,s}}{2\Delta\omega\biggl[1 
  + 
  \frac{\alpha L +  G n_{th}}{\alpha L -G}\Bigl(e^{\alpha L -G}-1\Bigr)\biggr]}. \label{SNR20}
\end{align}
For simplicity, we assumed that the Brillouin acoustic mode frequency, $\Omega_m$, matches the frequency difference of the pump and Stokes, $\Omega$, meaning $\Omega-\Omega_m=0$. Additionally, we set $z=L$, where $L$ is the length of the amplifier and $G=G_BPL$ is the single pass gain. We now use Eq. \ref{SNRratio} to calculated the noise factor in the NPP approximation
\begin{align}\label{NFsim}
 F =& 1 + \frac{\alpha L +G n_{th}}{\alpha L -G}\bigl(e^{\alpha L -G}-1\bigr).
\end{align}
If Brillouin amplification is large with relatively low optical loss, $G\gg \alpha L$, then the noise factor simplifies to 
\begin{equation}\label{NFthermal}
F\approx 1+n_{th}.
\end{equation}
This result agrees with previous results for the NPP approximation when Brillouin gain is large \citep{olsson1987characteristics,desurvire2002erbium,otterstrom2020shaping}.

\section{Noise factor including phonon propagation}\label{sec:Full}
\begin{figure}[h!]
\includegraphics[width=8.5cm]{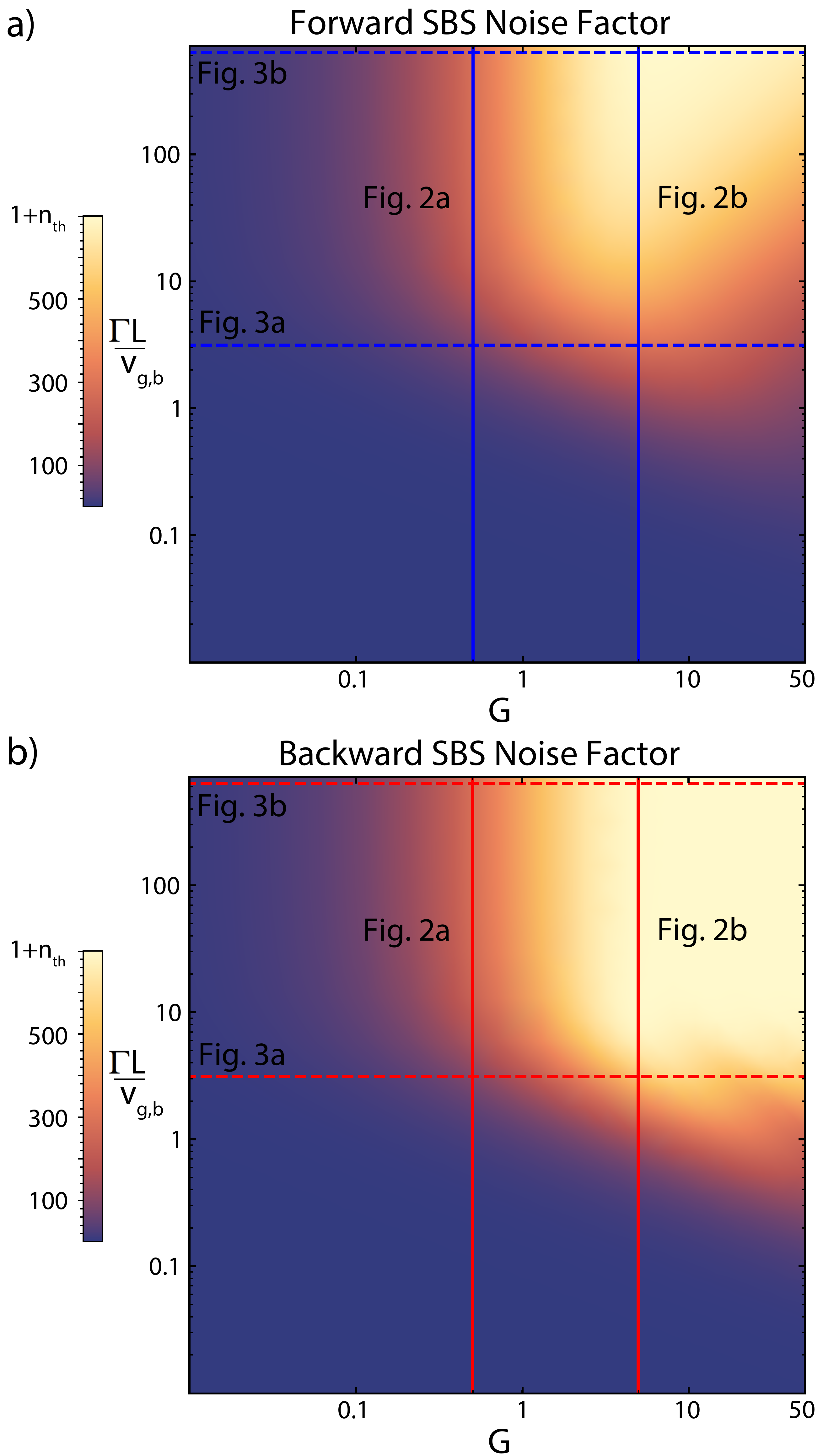}
\caption{\label{fig:Density}The noise factor of a forward intermodal SBS optical amplifier (a), Eq. \ref{NFfullmain}, and a backward Brillouin optical amplifier (b), Eq. \ref{NFbackfull}, for various values of $\Gamma L/v_{g,b}$ and $G$. This plot assumes no optical loss, $\alpha L=0$.  For the parameters chosen, $1+n_{th}=699.782$. In the lightest regions, we can see that while the noise factor approximation of $1+n_{th}$ holds true for many parameter combinations, but there are significant regions where the more complex models are necessary. At large $G$, there is a significant divergence in the noise factor between the forward and backward cases, seen clearly bisected by the Fig. \ref{fig:GB}a dashed lines. This is due to the forward intermodal Brillouin scattering interacting with coherent phonons at relatively larger phonon decay rates ($\Gamma L/v_{g,b}$). Figs. \ref{fig:Gamma} and \ref{fig:GB} are line plots of the noise factor with the path of those plots indicated on these density plots.}
\end{figure}

Further advancements in the field have produced systems\citep{bahl2012observation,kim2015non,dong2015brillouin,kharel2019high,kharel2022multimode,yoon2023simultaneous,xu2023strong,otterstrom2023modulation} for which the phonon propagation can no longer be neglected. In other words, the NPP approximation is no longer sufficient, and a nonzero $v_{g,b}$ is necessary to capture the salient dynamics. In this regime, the relative directions between the pump, Stokes, and acoustic waves velocities are now important. When all three waves are traveling in the same direction, this is called forward SBS. Specifically, since we do not include anti-Stokes scattering in this paper, we are modeling intermodal forward SBS, which breaks the symmetry of the process, decoupling anti-Stokes and Stokes processes \citep{otterstrom2018silicon,kittlaus2017chip,espinel2017brillouin,kang2010all}. When the Stokes wave is traveling opposite of the pump and acoustic wave, this is backward SBS.  The full analytical derivation of the noise factor for forward intermodal SBS can be found in Appendix \ref{appendixfullForward} and Appendix \ref{appendixfullBackward} for backward SBS. The results of these derivations, evaluated at the amplifier output ($z=L$), are 
\begin{widetext}
\noindent Forward intermodal SBS:
\begin{align}
 &F = 1 + e^{1/2(\alpha L + \Gamma L/v_{g,b})}\bigl[S_{A_{s,3}}(L)+S_{A_{s,4}}(L)\bigr]\Biggl|\cosh\Biggl(\frac{1}{2}\sqrt{\frac{\Gamma L}{v_{g,b}}G+\frac{1}{4}\biggl(\alpha L-\frac{\Gamma L}{v_{g,b}}\biggr)^2}\Biggr) \nonumber\\&+  
  \frac{\frac{1}{2}\bigl(\frac{\Gamma L}{v_{g,b}}-\alpha L\bigr)}{\sqrt{\frac{\Gamma L}{v_{g,b}}G+\frac{1}{4}\bigl(\alpha L-\frac{\Gamma L}{v_{g,b}}\bigr)^2}} \sinh\Biggl(\frac{1}{2}\sqrt{\frac{\Gamma L}{v_{g,b}}G+\frac{1}{4}\biggl(\alpha L-\frac{\Gamma L}{v_{g,b}}\biggr)^2}\Biggr)\Biggr|^{-2}\label{NFfullmain}
\end{align}
Backward SBS:
\begin{align}
F&=1+\frac{4}{\bigl|(\alpha L+\frac{\Gamma L}{v_{g,b}})^2-4G\frac{\Gamma L}{v_{g,b}}\bigr|}\nonumber\\ &\times\int_0^1 d\mathcal{Z} \biggl[ \biggl(\frac{\Gamma L}{ v_{g,b}}\biggr)^2 \frac{G n_{th}}{4}|e^{-L\lambda_+ \mathcal{Z}}-e^{-L\lambda_- \mathcal{Z}}|^2+\alpha L\Bigl|\Bigl(\frac{\Gamma L}{2 v_{g,b}}-L\lambda_+ \Bigr)e^{-L\lambda_+ \mathcal{Z}}-\Bigl(\frac{\Gamma L}{2 v_{g,b}}-L\lambda_- \Bigr)e^{-L\lambda_- \mathcal{Z}}\Bigr|^2\biggr]\label{NFbackfull}
\end{align}
where  $L\lambda_\pm = \frac{1}{4}\bigl(\Gamma L/v_{g,b}-\alpha L\bigr) \pm  \frac{1}{4}\sqrt{\bigl(\Gamma L/v_{g,b}+\alpha L\bigr)^2-4 G\frac{\Gamma L}{v_{g,b}}}$ and $\mathcal{Z}=z_1/L$ is the nondimensionalized spacial variable. The terms $S_{A_{s,3}}(L)$ and $S_{A_{s,4}}(L)$ are spectral densities quantifying the impact of the Langevin forces due to the phonons and photons, respectively, for forward intermodal SBS, given by
\begin{align}
 S_{A_{s,3}}(L)= &\biggl(\frac{\Gamma L}{v_{g,b}}\biggr)^2 \frac{n_{th} G}
  {\biggl|\frac{\Gamma L}{v_{g,b}}G+\frac{1}{4}\bigl(\alpha L-\frac{\Gamma L}{v_{g,b}}\bigr)^2\biggr|} \nonumber\\&\times\int_0^1 d\mathcal{Z} \ e^{-1/2(\alpha L + \Gamma L/v_{g,b})\mathcal{Z}}\Biggl|\sinh\Biggl(\frac{1}{2}\sqrt{\frac{\Gamma L}{v_{g,b}}G+\frac{1}{4}\biggl(\alpha L-\frac{\Gamma L}{v_{g,b}}\biggr)^2}\mathcal{Z}\Biggr)\Biggr|^{2}\label{As3PSDmain}
\end{align}
\begin{align} 
 S_{A_{s,4}}(L)=&\alpha L  \int_0^1 d\mathcal{Z} \  
  e^{-1/2(\alpha L + \Gamma L/v_{g,b})\mathcal{Z}}\Biggl|\cosh\Biggl(\frac{1}{2}\sqrt{\frac{\Gamma L}{v_{g,b}}G+\frac{1}{4}\biggl(\alpha L-\frac{\Gamma L}{v_{g,b}}\biggr)^2}\mathcal{Z}\Biggr) \nonumber\\&+  
  \frac{\frac{1}{2}\bigl(\frac{\Gamma L}{v_{g,b}}-\alpha L\bigr)}{\sqrt{\frac{\Gamma L}{v_{g,b}}G+\frac{1}{4}\bigl(\alpha L-\frac{\Gamma L}{v_{g,b}}\bigr)^2}} \sinh\Biggl(\frac{1}{2}\sqrt{\frac{\Gamma L}{v_{g,b}}G+\frac{1}{4}\biggl(\alpha L-\frac{\Gamma L}{v_{g,b}}\biggr)^2}\mathcal{Z}\Biggr)\Biggr|^{2}\label{As4PSDmain}. 
\end{align}
\end{widetext} 
\indent In the next subsections, we will analyze Eqs. \ref{NFfullmain} and \ref{NFbackfull}, as well as demonstrate the regimes where the full dynamics converge with both the NPP approximation (Eq. \ref{NFsim}) and $F\approx 1+n_{th}$ (Eq. \ref{NFthermal}), and where these approximations are no longer sufficient. To do this, we will examine three nondimensionalized variables $\Gamma L/v_{g,b}$, $G$, and $\alpha L$ to capture the effects of phonon decay, Brillouin gain, and photon decay, respectively. For the first two subsections, we assume a lossless optical amplifier with $\alpha L=0$. In third subsection, we examine the impact of including optical loss on the noise factor. To start, we present density plots of the forward intermodal SBS noise figure (see Fig. \ref{fig:Density}a), Eq. \ref{NFfullmain}, and backward SBS scattering (see Fig. \ref{fig:Density}b), Eq. \ref{NFbackfull}, for values of $\Gamma L/v_{g,b}$ and $G$, with no optical loss ($\alpha L=0$). In the darker regions of this figure, we can see that the approximate noise factor given by Eq. \ref{NFthermal} ($F\approx 1+n_{th}$) does not accurately capture the amplifier performance, necessitating the use of our full dynamical models, Eqs. \ref{NFfullmain} and \ref{NFbackfull}. We have assumed a system temperature of 300 K and a phonon frequency of $\Omega/(2 \pi)= 8.78$GHz, resulting in $n_{th}=698.782$ for the calculations in this paper.

\subsection{Noise factor as a function of phonon decay} 
\begin{figure}[h!]
\includegraphics[width=8.5cm]{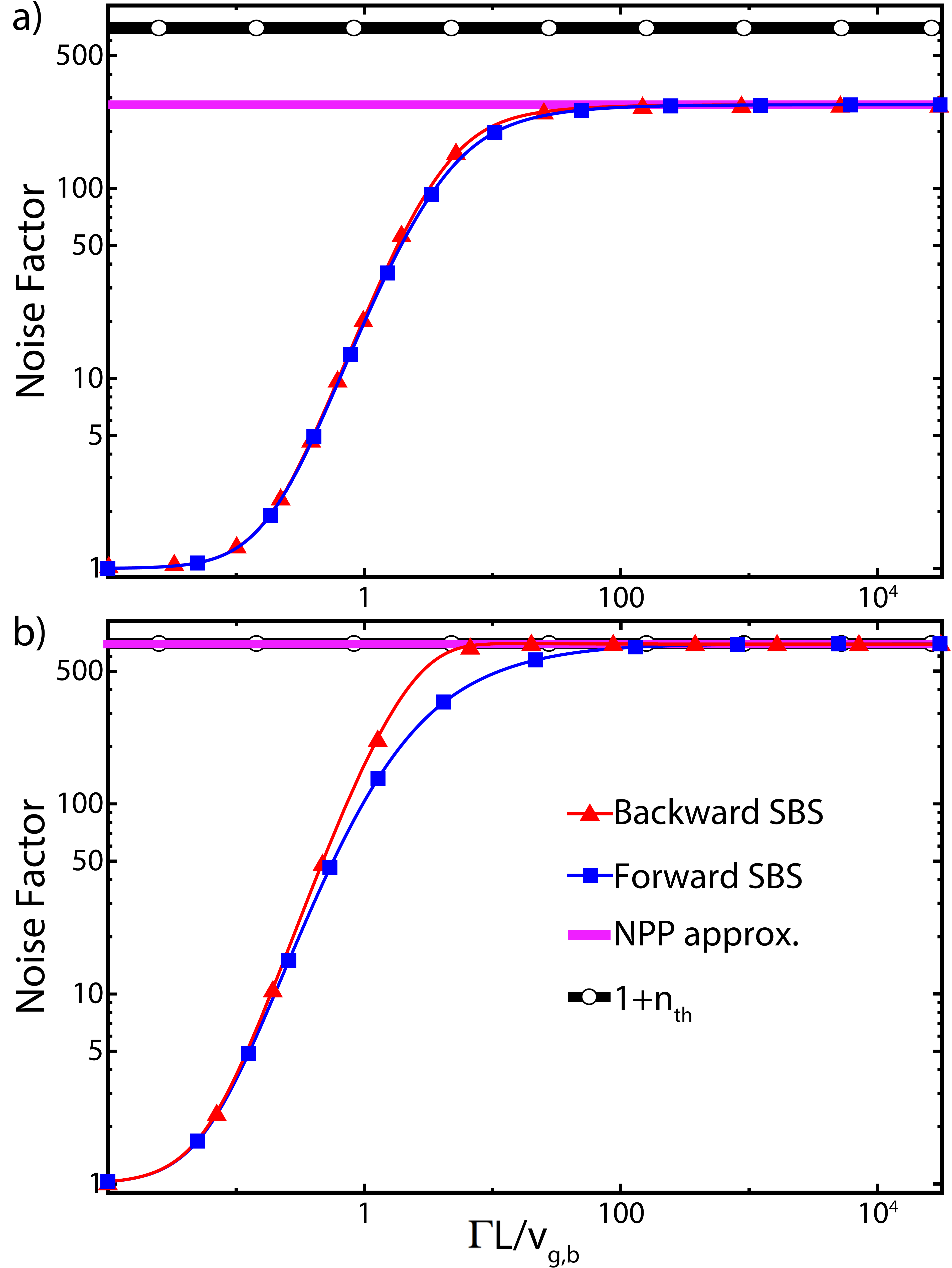}
\caption{\label{fig:Gamma}Plot of a Brillouin optical amplifier's noise factor a function of the nondimensionalized phonon decay $\Gamma L/v_{g,b}$. In plot (a) we have assumed a single pass Brillouin gain of $G=0.5$ and in plot (b), $G=5$. Additionally, it is assumed that there is no optical loss ($\alpha=0$).  For context, if $L=0.01$ m and $v_{g,b}=$ 3000 m/s, then this would be a plot of phonon decay rates ($\Gamma$) from 3 kHz to 10 GHz. In each plot, we have calculated the noise factor with the chosen parameters using the thermal approximation, $1+n_{th}$ (black, open circles), the non-propagating phonon (NPP) approximation (magenta, Eq. \ref{NFsim}), forward intermodal SBS (blue squares, Eq.\ref{NFfullmain}), and backward SBS (red triangles, Eq. \ref{NFbackfull}).}
\end{figure} 
The SBS amplification process takes place in a gain window with a width determined by the phonon decay rate. Emerging Brillouin systems can have phonon decay rates small enough that the NPP approximation is no longer valid, namely when the phonon mean free path is comparable to the system length. We plot the noise factor as a function of nondimensionalized phonon decay, $\Gamma L/v_{g,b}$, in Fig. \ref{fig:Gamma}, in order to quantify at what ranges the NPP approximation breaks down. In Fig. \ref{fig:Gamma}a, we assume a single pass gain of $G=0.5$. We plot a forward intermodal SBS amplifier (Eq. \ref{NFfullmain}) in the blue (squares) and a backward SBS amplifier (Eq. \ref{NFbackfull}) in red (triangles). We observe that at large phonon decay rates, $\Gamma L/v_{g,b}\gtrsim 100$, the NPP approximation (magenta, Eq. \ref{NFsim}) is consistent with the full dynamical model, as expected. In Fig. \ref{fig:Gamma}b, we plot the same equations with a larger single pass gain of $G=5$. We see that in this case, the forward intermodal, backward, and NPP approximation plots converge to the thermal result of $F\approx 1+n_{th}$ (black, open circles, Eq. \ref{NFthermal}), at large phonon decay rates, due to the large single pass gain, $G\gg\alpha L$. However, in both plots it is evident the importance of accounting for phonon propagation.

\subsection{Noise factor as a function of single pass gain } 
\begin{figure}[h!]
\includegraphics[width=8.5cm]{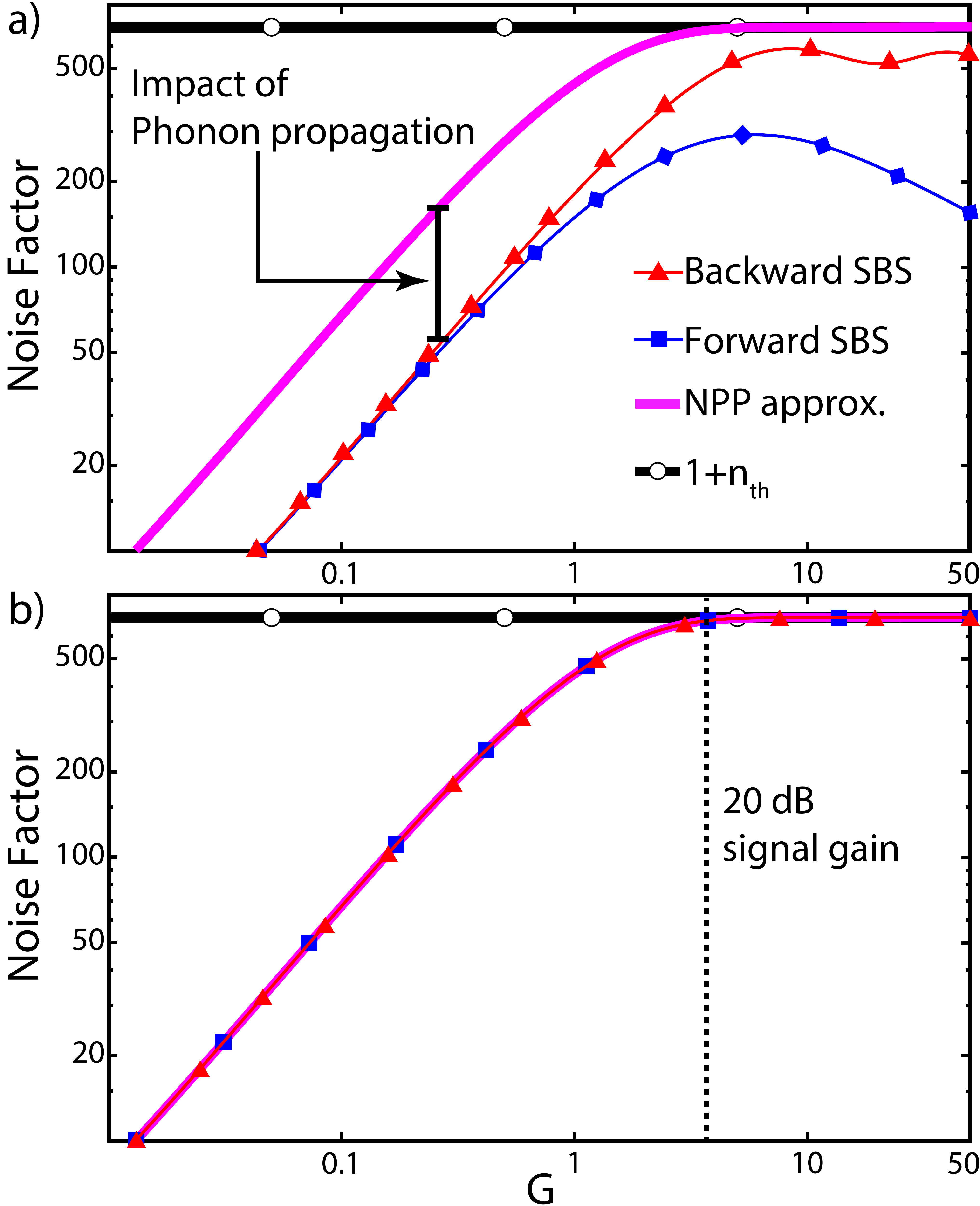}
\caption{\label{fig:GB}Plot of a Brillouin optical amplifier's noise factor a function of the single pass gain $G$. In plot (a) we have assumed a nondimensionalized phonon decay of $\Gamma L/v_{g,b}=2 \pi\times 0.5$ and in plot (b), $\Gamma L/v_{g,b}=2 \pi \times 100$. It is assumed that there is no optical loss ($\alpha=0$). In each plot, we have calculated the noise factor with the chosen parameters using the thermal approximation, $1+n_{th}$ (black, open circles), the non-propagating phonon (NPP) approximation (magenta, Eq. \ref{NFsim}), forward intermodal SBS (blue squares, Eq.\ref{NFfullmain}), and backward SBS (red triangles, Eq. \ref{NFbackfull}). In plot (b) the Stokes signal experiences exponential growth according to $exp\{G/2\}$, so for reader context we indicate the single pass gain according to 20 dB. In plot (a), the backward SBS case has signal behavior at values of $G>5$ that begins to violate the undepleted pump approximation. See appendix \ref{appendixamplitude} for an in depth discussion on this behavior.}
\end{figure}

The second variable of interest is the single pass gain, $G=G_BPL$. In Fig. \ref{fig:GB} we plot the noise factor as a function of this variable. In Fig. \ref{fig:GB}a, the phonon decay term is assumed to be $\Gamma L/v_{g,b}=2\pi \ 0.5$. If  $L=0.01$ m and $v_{g,b}=$ 3000 m/s, then this would correspond to a phonon decay of $\Gamma=2\pi \ 1.5 \times 10^5$ Hz. In this case, when the phonon dissipation is low and consequently the phonon mean-free path is large, the full dynamics exhibit a significant correction to the result from both the NPP approximation (Eq. \ref{NFsim}) and the inferred noise figure limit (Eq. \ref{NFthermal}), as shown Fig. \ref{fig:GB}a. The reason for the significant decrease in the noise factor is that the injected signal and coherent phonon fields grow spatially at an exponential rate across the entire Brillouin interaction region. By contrast, thermal phonons are generated throughout the length of the amplifier, meaning that the associated noise only grows over a partial interaction length. In other words, in the case of significant phonon propagation we observe that the optomechanical backaction experienced by the thermal fluctuations is fundamentally distinct from that experienced by the coherent signal. Furthermore, the noise factor in forward Brillouin scattering (blue squares) exhibits a notable decrease at large $G$. At these values, the rate of growth of the noise starts to decrease relative to the rate of growth of the signal, meaning an overall increase in $SNR_2$. 
\newline
\indent The noise factor of backward SBS including phonon propagation also exhibits a dramatic departure from the conventional NPP theory. The unique boundary conditions required by backward SBS arise from the fact that initial conditions for the Stokes and phonon fields are defined only on the input and output ends, respectively. This creates a series of pseudo-gain resonances in the coherent exchange of energy between optical and acoustic domains, marked by divergences in the signal and noise amplitudes. While the noise factor remains well behaved, there exists a slight phase difference between signal and noise, causing the oscillation in the noise factor (red triangles) in Fig. \ref{fig:GB}a. This result, along with the forward SBS case, demonstrates that under the proper conditions, Brillouin amplifiers are significantly less noisy than previously assumed by the NPP approximation. 
Furthermore, when applying these models in systems with large G, it is important to consider the validity of the undepleted pump approximation. The undepleted (or stiff) pump approximation can break down in the case of sufficiently large Brillouin gain and/or Stokes input, or a combination thereof.  These considerations require close examination in the case of backward SBS systems with significant phonon propagation producing the pseudo-gain resonances. As shown in Appendix \ref{appendixamplitude}, however, this divergent behavior is an artifact of the undepleted pump approximation (providing infinite gain for the Stokes and phonon fields), and we recover physical amplitude solutions when accounting for pump depletion.
\newline
\indent In contrast, we choose a phonon decay that satisfies the NPP approximation in Fig. \ref{fig:GB}b, $\Gamma L/v_{g,b}=2\pi \ 100$, which is $\Gamma=2\pi \ 10^{7}$ Hz, assuming the previous values. We see that the full analytical models converge to the NPP approximation, as expected. Additionally, the models converge to $F\approx 1+n_{th}$ (Eq. \ref{NFthermal}), once $G\gtrsim 5$. 

\subsection{Noise factor as a function of optical loss} 
\begin{figure}[h!]
\includegraphics[width=8.5cm]{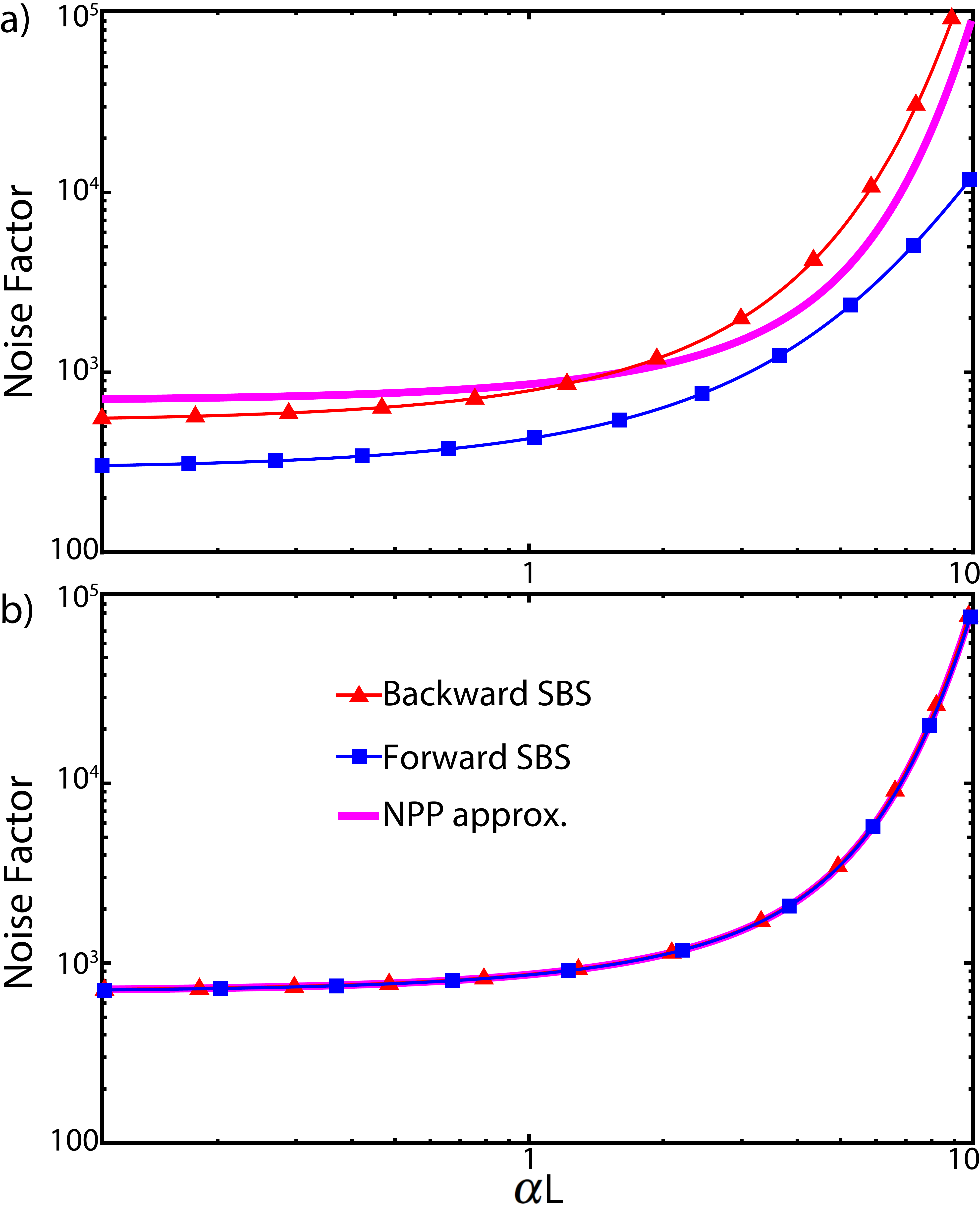}
\caption{\label{fig:alpha}Plot of a Brillouin optical amplifier's noise factor as a function of the nondimensionalized photon decay $\alpha L$. Forward intermodal SBS (blue squares, Eq. \ref{NFfullmain}, backward SBS (red triangles, Eq. \ref{NFbackfull}, and the non-propagating phonon (NPP) approximation (magenta, Eq. \ref{NFsim}). As we expect, the noise factor rapidly grows as the optical decay of the photons become the dominant effect in our optical amplifiers.  In part (a), the nondimensionalized phonon decay is $\Gamma L/v_{g,b}=2\pi \ 0.5$. In part (b), the nondimensionalized phonon decay is $\Gamma L/v_{g,b}=2\pi \ 100$. Both parts assume a single pass gain of $G=5$.  As expected, in part (b) the three plots converge due to the large phonon decay, satisfying the NPP approximation.}
\end{figure}

While optical loss is not the dominant impact on the noise factor, it is important to account for effect of this variable for less optically efficient systems. In Fig. \ref{fig:alpha}, we plot the noise factor as a function of nondimensionalized optical loss, $\alpha L$.  We see that the performance of Brillouin amplifiers decays rapidly as optical increases.  In Fig. \ref{fig:alpha}a, we plot forward intermodal SBS (blue squares, Eq. \ref{NFfullmain}), backward SBS (red triangles, Eq. \ref{NFbackfull}, and the NPP approximation (magenta, Eq. \ref{NFsim} with a phonon decay of $\Gamma L/v_{g,b}=2\pi \ 0.5$ and single pass gain of $G=5$. We see that the NPP approximation is not sufficient in this regime. In Fig. \ref{fig:alpha}b, we plot the same equations with a phonon decay of $\Gamma L/v_{g,b}=2\pi \ 100$ and single pass gain of $G=5$. In this case, we see the three plots converge, as expected.

\section{Discussion} \label{sec:discuss}

In this paper we have calculated a Brillouin amplifier's noise factor using a traveling wave-based Hamiltonian model. Our model accounts for the spatial propagation of the phonons and whether the amplifier is operating in the backward or forward intermodal SBS configuration. Under the NPP approximation, valid when phonon decay is large, our model converges to the previously measured and inferred noise factor result of $F\approx 1+n_{th}$. We demonstrate the behavior of our model as a function of several nondimensionalized variables that account for the impact of phonon decay, Brillouin gain, and photon decay. Importantly, our model indicates where the NPP approximation does not accurately capture the dynamics of a Brillouin amplifier by severely overestimating the noise contribution, as the thermal mechanical noise experiences a much shorter interaction length than the injected signal. 
\newline
\indent Our model will provide a valuable tool for a growing range of Brillouin systems whose geometry and material composition results in parameters that exist outside the NPP approximation\citep{bahl2012observation,kim2015non,dong2015brillouin,kharel2019high,kharel2022multimode,yoon2023simultaneous,xu2023strong,otterstrom2023modulation}. In particular, Ref. \onlinecite{otterstrom2023modulation} demonstrates the possibility of tunable phonon propagation using acoustoelectrics, enabling optomechanical dynamics well past the regime accurately modeled by the NPP approximation. Our full dynamical models offer a much deeper understanding of the noise properties in these rapidly evolving systems.

\section*{Acknowledgments}

This material is based upon work supported by the U.S. Department of Energy, Office of Science, Office of Workforce Development for Teachers and Scientists, Office of Science Graduate Student Research (SCGSR) program. The SCGSR program is administered by the Oak Ridge Institute for Science and Education (ORISE) for the DOE. ORISE is managed by ORAU under contract number DE-SC0014664. Sandia National Laboratories is a multi-mission laboratory managed and operated by National Technology \& Engineering Solutions of Sandia, LLC (NTESS), a wholly owned subsidiary of Honeywell International Inc., for the U.S. Department of Energy’s National Nuclear Security Administration (DOE/NNSA) under contract DE-NA0003525. This written work is authored by an employee of NTESS. The employee, not NTESS, owns the right, title and interest in and to the written work and is responsible for its contents. This paper describes objective technical results and analysis. Any subjective views or opinions that might be expressed in the paper do not necessarily represent the views of the U.S. Department of Energy or the United States Government. The publisher acknowledges that the U.S. Government retains a non-exclusive, paid-up, irrevocable, world-wide license to publish or reproduce the published form of this written work or allow others to do so, for U.S. Government purposes. The DOE will provide public access to results of federally sponsored research in accordance with the DOE Public Access Plan. The authors have no conflicts to disclose.
\newline

\appendix

\section{Full phonon dynamics: forward Brillouin scattering}\label{appendixfullForward}
To capture the dynamics of a forward intermodal SBS optical amplifier outside the NPP approximation and compare to Eqs. \ref{NFsim} and \ref{NFthermal}, $v_{g,b}$ is much greater than zero. We transform  Eqs. \ref{Beom} and \ref{Aseom} into the Fourier domain $\{F(\omega) = \int_{-\infty}^\infty dt f(t) e^{i \omega t} \}$ in time and the Laplace domain $\{F(s) = \int_0^\infty dz f(z) e^{-s z} \}$ in space, where $s$ is a complex-valued frequency that has a complex conjugate $s^*$. We select the Laplace transform for space because we want to account for the initial conditions of our system at $z=0$, noting that $\mathfrak{L}\{\partial f(z)/\partial z\} = s F(s) - f(z=0)$. The transformed equations of motion are
\begin{align} 
 &\Bigl[i(\Omega - \Omega_m -\omega) + \frac{\Gamma}{2} + v_{g,b} s^*\Bigr]\overline{B}
  (s^*,\omega)\nonumber\\ &= v_{g,b} \overline{B}(z=0,\omega) - i g_0^* 
  \overline{A}_p \overline{A}_s^\dagger(s^*,\omega) + \eta(s^*,
  \omega)\label{BTreomfull} \\ 
 &\Bigl(-i\omega + \frac{\alpha v_{g,s}}{2} + v_{g,s} s\Bigr)\overline{A}_s(s,\omega)
  \nonumber\\ &= v_{g,s} \overline{A}_s(z=0,\omega) - i g_0^* \overline{A}_p 
  \overline{B}^\dagger(s,\omega) + \xi_s(s,\omega), \label{AsTreomfull}
\end{align}
where $\overline{A}_s(z=0,\omega)$ and $\overline{B}(z=0,\omega)$ are the input noise field for the Stokes and phonon fields, respectively. Then we solve for $\overline{B}$ in Eq. \ref{BTreomfull} and taking the Hermitian conjugate, giving
\begin{equation}\label{BdagC1}
 \overline{B}^\dagger(s,\omega) = \frac{v_{g,b} \overline{B} 
  ^\dagger(z=0,\omega) + i g_0 A_p^\dagger \overline{A}_s(s,\omega) + 
  \eta^\dagger(s,\omega)}{v_{g,b} s + {\chi^*_B}^{-1}},
\end{equation}
where the phonon susceptibility is $\chi_B=\bigl[i(\Omega - \Omega_m -\omega) + \Gamma/2\bigr]^{-1}$. Equation \ref{BdagC1} is inserted into Eq. \ref{AsTreomfull}, which results in the decoupled Stokes' field amplitude,
\begin{align}\label{AsC2}
 \overline{A}_s(s,\omega) &= \frac{v_{g,s}(v_{g,b} s + {\chi^*_B}^{-1})}
  {v_{g,s}v_{g,b} s^2 + \chi^*_C s + \chi^*_A} \overline{A}_  
  s(z=0,\omega)\nonumber \\ &-\frac{i g_0^* \overline{A}_p v_{g,b}}
  {v_{g,s}v_{g,b} s^2 + \chi^*_C s + \chi^*_A} \overline{B} 
  ^\dagger(z=0,\omega)\nonumber \\ &-\frac{i g_0^* \overline{A}_p}
  {v_{g,s}v_{g,b} s^2 + \chi^*_C s + \chi^*_A} 
  \eta^\dagger(s,\omega)\nonumber \\ &+\frac{v_{g,b} s + {\chi^*_B}^{-1}}
  {v_{g,s}v_{g,b} s^2 + \chi^*_C s + \chi^*_A} \xi_  
  s(s,\omega)
\end{align}
where $\chi_A= \chi_B^{-1}(i\omega + \alpha v_{g,s}/2)-v_{g,s} G_B P \Gamma/4$ and $\chi_C= (\alpha v_{g,s}/2)v_{g,b} + i \omega v_{g,b} + v_{g,s} {\chi_B}^{-1}$. 
\newline Taking the inverse Laplace transform and noting that $\mathfrak{L}^{-1}\{F(s)H(s)\} = f(z)\star h(z)$, where $\star$ denotes convolution starting from $z_1=0$. The factors of s for each term in the equation above are manipulated into well known inverse Laplace transforms. The Stokes field amplitude in the space domain is
\begin{widetext}
\begin{align}
 &\overline{A}_s(z,\omega) = \overline{A}_{s,1}(z,\omega)+
  \overline{A}_{s,2}(z,\omega)+\overline{A}_{s,3}(z,\omega)+\overline{A}_{s,
  4}(z,\omega),\label{AsC2zw} \\
 &\overline{A}_{s,1}(z,\omega)= e^{- \chi^*_C z/(2 v_{g,s}
  v_{g,b})}\Bigl[\cos(P_1 z) + \frac{P_2}{P_1}
  \sin(P_1 z)\Bigr]\overline{A}_s(z=0,\omega),\label{As1}\\
 &\overline{A}_{s,2}(z,\omega)=-e^{- \chi^*_C z/(2 v_{g,s}
  v_{g,b})}\frac{i g_0^* \overline{A}_p}{v_{g,s}P_1}\sin(P_1
  z)\overline{B}^\dagger(z=0,\omega),\label{As2}\\
 &\overline{A}_{s,3}(z,\omega)=\frac{i g_0^* \overline{A}_p}{v_{g,s}v_{g,b}
  P_1} \int_0^z dz_1 e^{- \chi^*_C (z-
  z_1)/(2 v_{g,s}v_{g,b})}\sin\bigl[P_1(z-z_1)\bigr]\eta^\dagger(z_1,\omega),
  \label{As3}\\
 &\overline{A}_{s,4}(z,\omega)=\frac{1}{v_{g,s}} \int_0^z dz_1 e^{-
  \chi^*_C (z-z_1)/(2 v_{g,s}v_{g,b})}\Bigl\{\cos\bigl[P_1 (z-z_1)\bigr] +
  \frac{P_2}{P_1} \sin\bigr[P_1 (z-z_1)\bigr]\Bigr\}\xi_s(z_1,
  \omega),\label{As4}
\end{align}
and $P_1=\sqrt{\frac{\chi^*_A}{v_{g,s} v_{g,b}}-\Bigl(\frac{\chi^*_c}{2v_{g,s} v_{g,b}}\Bigr)^2}$ and $P_2=\frac{1}{v_{g,b}\chi^*_B}-\frac{\chi^*_c}{2v_{g,s} v_{g,b}}$. Give that the origins of the individual amplitudes in Eq. \ref{AsC2zw} are physically distinct, we assume that they are uncorrelated with each other. Therefore, the spectral density of the Stokes field amplitude, $S_{A_s}(z,\omega)$, is
\begin{align}\label{AsPSD}
S_{A_s}(z,\omega) = S_{A_{s,1}}(z,\omega) + S_{A_{s,2}}(z,\omega) + S_{A_{s,3}}(z,\omega) + S_{A_{s,4}}(z,\omega) 
\end{align}
where
\begin{align}
 &S_{A_{s,1}}(z,\omega)= e^{- Re[\chi^*_C] z/( v_{g,s}
  v_{g,b})}\Bigl|\cos(P_1 z) + \frac{P_2}{P_1} 
   \sin(P_1 z)\Bigr|^2 \Bigl[S_{A_s}^{C}(0,\omega) + S_{A_s}^{N}(0,\omega)\Bigr]
  \label{As1PSD}\\
 &S_{A_{s,2}}(z,\omega)= e^{-Re[\chi^*_C] z/(v_{g,s}
  v_{g,b})}\frac{G_B P \Gamma}{4 v_{g,s}|P_1|^2}
  \bigl|\sin(P_1 z)\bigr|^2 S_{B}^{in}(0,\omega)\label{As2PSD}\\
 &S_{A_{s,3}}(z,\omega)= \frac{\Gamma^2 n_{th} G_B P}
  {4 v_{g,s}v_{g,b}^2 |P_1|^2} \int_0^z dz_1 e^{- 
  Re[\chi^*_C] (z-z_1)/(v_{g,s}v_{g,b})}\Bigl|\sin\bigl[P_1 (z-z_1)\bigr]\Bigr|
  ^2\label{As3PSD}\\
 &S_{A_{s,4}}(z,\omega)=\frac{\alpha}{v_{g,s}} \int_0^z dz_1 
  e^{-Re[\chi^*_C] (z-z_1)/(v_{g,s}v_{g,b})}\Bigl|\cos\bigl[P_1 (z- 
  z_1)\bigr] +\frac{P_2}{P_1} \sin\bigl[P_1 (z-z_1)\bigr]\Bigr|
  ^2\label{As4PSD}. 
\end{align}
\newline
\indent We note that the $S_{A_s}^{C}(0,\omega)$ term is the amplified injected signal while the rest of the terms in these equations are the noise of the amplification process. Additionally, we make the assumption that the injected phonon noise PSD is zero, $S_{B}^{in}(0,\omega)=0$. This is the case when the material coupling the pump light into the optical amplifier differ such that the phonons generated in the coupling material fall outside of the phase matching conditions of the Brillouin optical amplifier, contributing no injected phonon noise. Considering the noise factor for bandwidths ($\Delta\omega$) that are much smaller than $\Gamma$, we use Eqs. \ref{As1PSD}, \ref{As3PSD}, and \ref{As4PSD} and the output signal-to-noise ratio is
\begin{align}\label{SNR2}
 &SNR_2 = \frac{|A_s^{in}|^2  e^{- Re[\chi^*_C] z/( v_{g,s}
  v_{g,b})}\bigl|\cos(P_1 z) + \frac{P_2}{P_1} 
   \sin(P_1 z)\bigr|^2}{2\Delta\omega\Bigl[\frac{1}{v_{g,s}}e^{- 
  Re[\chi^*_C] z/( v_{g,s}v_{g,b})}\bigl|\cos(P_1 z) + \frac{P_2}{P_1} 
   \sin(P_1 z)\bigr|^2+S_{A_{s,3}}(z)+S_{A_{s,4}}(z)\Bigr]}
\end{align}
Using Eq.\ref{SNRratio}, the noise factor ($F$) is
\begin{align}\label{NFfull}
 &F = 1 + \frac{v_{g,s}\bigl[S_{A_{s,3}}(z)+S_{A_{s,4}}(z)\bigr]}{e^{-  
  Re[\chi^*_C] z/( v_{g,s}v_{g,b})}\bigl|\cos(P_1 z) + \frac{P_2}{P_1} 
   \sin(P_1 z)\bigr|^2},
\end{align}
matching the result in main body, Eq. \ref{NFfullmain}.
\end{widetext}

\section{Full phonon dynamics: backward Brillouin scattering}\label{appendixfullBackward}
In the case of backward Brillouin scattering, a different approach is necessary since the initial conditions of the pump and acoustic waves begin opposite of the Stokes wave.  A Laplace transform is no longer appropriate in this case, and we will use a Green's function approach. To start, we slightly modify our equations of motion so that the velocities have the proper sign. We apply a Fourier transform in time as described in the main body, which results in
\begin{align} 
 \frac{\partial}{\partial z}\overline{B}(z,\omega)-\Lambda_B\overline{B}(z,\omega)
    - \frac{1}{v_{g,b}}&i g_0^*\overline{A}_p \overline{A}_s^\dagger(z,\omega) \nonumber\\ & =- \frac{1}{v_{g,b}}\eta(z,\omega)
    \label{BeomBacksimp} 
 \end{align}
\begin{align} 
 \frac{\partial}{\partial z}\overline{A}_s(z,\omega)-\Lambda_S \overline{A}_s(z,\omega) + \frac{1}{v_{g,s}}& i g_0^* \overline{A}_p \overline{B}^\dagger(z,\omega)\nonumber\\ & = \frac{1}{v_{g,s}}\xi_s(z,\omega). \label{AseomBacksimp} 
\end{align}
where $\Lambda_b = 1/v_{g,b}\bigl[i (\Omega_m - \Omega-\omega)+\frac{\Gamma}{2}\bigr]= 1/v_{g,b}{\chi_B}^{-1}$ and $\Lambda_S = -1/v_{g,s}\bigl(-i \omega+\frac{\gamma_s}{2}\bigr)$. Operating with $\partial z - \Lambda_b^*$ on Eq. \ref{AseomBacksimp} and eliminating $\overline{B}^\dagger(z,\omega)$ using Eq. \ref{BeomBacksimp}, we obtain	
\begin{align}
  &\Biggl[\frac{\partial^2}{\partial z^2}-(\Lambda_S+\Lambda^*_b)\frac{\partial}{\partial z}
   +\biggl(\Lambda_S\Lambda^*_b+\frac{|g_0|^2|\overline{A}_p|^2}{v_{g,s}v_{g,b}}\biggr)\Biggr]\overline{A}
   _s(z,\omega)\nonumber\\
  &=f_s(z,\omega),
  \label{Backeomfinal}
\end{align}
where $f_s(z,\omega)=1/v_{g,s}(\partial z-\Lambda^*_b)\xi_s(z,\omega)+ig^*_0 \overline{A}_p/(v_{g,s}v_{g,b})\eta(z,\omega)$.

\subsection{Boundary conditions}
To properly quantify the boundary conditions of a backward SBS amplifier, we inject the seed Stokes light at z=0, so that $\overline{A}_s(0)$ is a specified quantity we control, and, at z=L, we use Eq. \ref{AseomBacksimp} to find the boundary condition,
\begin{align}\label{BackBC}
  \biggl[\frac{\partial}{\partial z}\overline{A}_s(z)-\Lambda_s&
  \overline{A}_s(z)\biggr]\biggl|_{z=L} \nonumber\\ &=  - \frac{1}{v_{g,s}}i g_0^* \overline{A}_p \overline{B}^\dagger(L) +  
  \frac{1}{v_{g,s}}\xi_s(L).
\end{align}
We make the assumption $\overline{B}^\dagger(L)=0$, due to the same justification listed in Appendix \ref{appendixfullForward} of no injected phonon noise.
\subsection{Green's function}
For our Stokes amplitude solution, we start with
\begin{equation}\label{ASdelta}
\overline{A}_s(z,\omega)=\int^L_0 dz' \delta(z-z') \overline{A}_s(z',\omega).
\end{equation}
Using the definition for the Stokes amplitude Green functions $\Bigl[\frac{\partial^2}{\partial z^2}-(\Lambda_S+\Lambda^*_b)\frac{\partial}{\partial z}
   +\Bigl(\Lambda_S\Lambda^*_b+\frac{|g_0|^2|\overline{A}_p|^2}{v_{g,s}v_{g,b}}\Bigr)\Bigr]G_A(z,z')=\delta (z-z')$, two repetitions of integration by parts, and the application of the boundary conditions, Eq. \ref{ASdelta} becomes  
\begin{align}\label{ASgreensol}
\overline{A}_s(z,\omega)=&-\overline{A}_s(0,\omega)\frac{\partial}{\partial z'}G_A(z,z')\Big |_{z'=0}^L\nonumber\\ &-\frac{1}{v_{g,s}}\big [\xi_s(L,\omega) - i g_0^* \overline{A}_p \overline{B}^\dagger(L,\omega)\big ]G_A(z,L) \nonumber\\ &+\int^L_0 dz' G_A(z,z') f_s(z',\omega).
\end{align}
To solve for an appropriate Green's function for our system, we break up the solution into two regions: $G_-(z,z')$ when $z<z'$ and $G_+(z,z')$ when $z>z'$. These regions have the forms
\begin{align}
&G_-(z,z')=A e^{\lambda_+(z-z')}+B e^{\lambda_-(z-z')}\label{Gminusgen}\\
&G_+(z,z')=C e^{\lambda_+(z-z')}+D e^{\lambda_-(z-z')}\label{Gplusgen},
\end{align}
where $\lambda_+$ and $\lambda_-$ are the eigenvalues of the homogeneous eigenvalue characteristic polynomial, $\lambda^2-(\Lambda_S+\Lambda^*_b)\lambda+\Bigl(\Lambda_S\Lambda^*_b+\frac{|g_0|^2|\overline{A}_p|^2}{v_{g,s}v_{g,b}}\Bigr)=0$. The eigenvalues are 
\begin{align}\label{eigen}
\lambda_\pm = \frac{1}{2}(\Lambda_S+\Lambda^*_b) \pm  \frac{1}{2}\sqrt{(\Lambda_S-\Lambda^*_b)^2-\frac{|g_0|^2|\overline{A}_p|^2}{v_{g,s}v_{g,b}}}
\end{align}
To specify the unknown coefficients in Eqs. \ref{Gminusgen} and \ref{Gplusgen}, we use the Green's function jump condition, meaning that $G_+$ and $G_-$ are continuous at $z=z'$, $\bigl[\partial z G_+(z,z')-\partial z G_-(z,z')\bigr]\big |_{z=z'}=1$, and the boundary conditions $G_A(z,0)=0$ and $\bigl[\frac{\partial}{\partial z}+\Lambda_b\bigr]G_A(z,z')\big |_{z'=L} =0$, giving
\begin{align}
&G_-(z,z')=A(z) e^{-\lambda_+z'}+B(z) e^{-\lambda_-z'}\label{Gminuspar}\\
&G_+(z,z')=A(z) e^{-\lambda_+z'}+B(z) e^{-\lambda_-z'}+\Psi(z-z')\label{Gpluspar},
\end{align} 
where
\begin{align}
&\Psi(z)=\frac{1}{\lambda_+-\lambda_-}\bigl[e^{\lambda_+ z}-e^{\lambda_- z}\bigr]\label{Psi}\\
&A(z)=-\frac{\Lambda_b^*-\lambda_-}{\Lambda_b^*-\lambda_-(\Lambda_b^*-\lambda_+)e^{-(\lambda_+-\lambda_-)L}}\Psi(z)\label{GA}\\&B(z)=\frac{(\Lambda_b^*-\lambda_+)e^{-(\lambda_+-\lambda_-)L}}{\Lambda_b^*-\lambda_-(\Lambda_b^*-\lambda_+)e^{-(\lambda_+-\lambda_-)L}}\Psi(z) \label{GB}.
\end{align}
Eqs. \ref{Gminuspar} and \ref{Gpluspar} can be stitched together with a Heaviside theta function, giving our system's Green's function
\begin{align}
&G_A(z,z')\nonumber\\ &=A(z) e^{-\lambda_+z'}+B(z) e^{-\lambda_-z'}+\Theta(z-z')\Psi(z-z').
\end{align}
We can plug this into Eq. \ref{ASgreensol}, giving our Stokes amplitude
\begin{align}\label{ASbackfull}
&\overline{A}_s(z,\omega)\nonumber\\&=-\overline{A}_s(0,\omega)\frac{\partial}{\partial z'}G_+(z,z')\Big |_{z'=0}^L\nonumber\\ &- i g_0^* \overline{A}_p \overline{B}^\dagger(L,\omega)G_-(z,L) \nonumber\\ &-\frac{1}{v_{g,s}}\int^L_0 dz' G_A(z,z') \Bigl[\Lambda_b^*\xi_s(z',\omega)+\frac{i g_0^* \overline{A}_p}{v_{g,b}}\eta^\dagger(z',\omega)\Bigr]\nonumber\\ &-\frac{1}{v_{g,s}}\int^L_0 dz'\biggl\{\nonumber\\ & \xi_s(z',\omega)\biggl[\frac{\partial}{\partial z}G_-(z,z')+\Theta(z-z')\frac{\partial}{\partial z}\Psi(z-z')\biggr]\biggr\}.
\end{align}
Using Eq. \ref{ASbackfull}, the noise factor of backward Brillouin scattering is calculated according to previous sections, leading to Eq. \ref{NFbackfull}.

\section{Stokes signal amplitude analysis}\label{appendixamplitude}
\begin{figure}[h!]
\includegraphics[width=8.5cm]{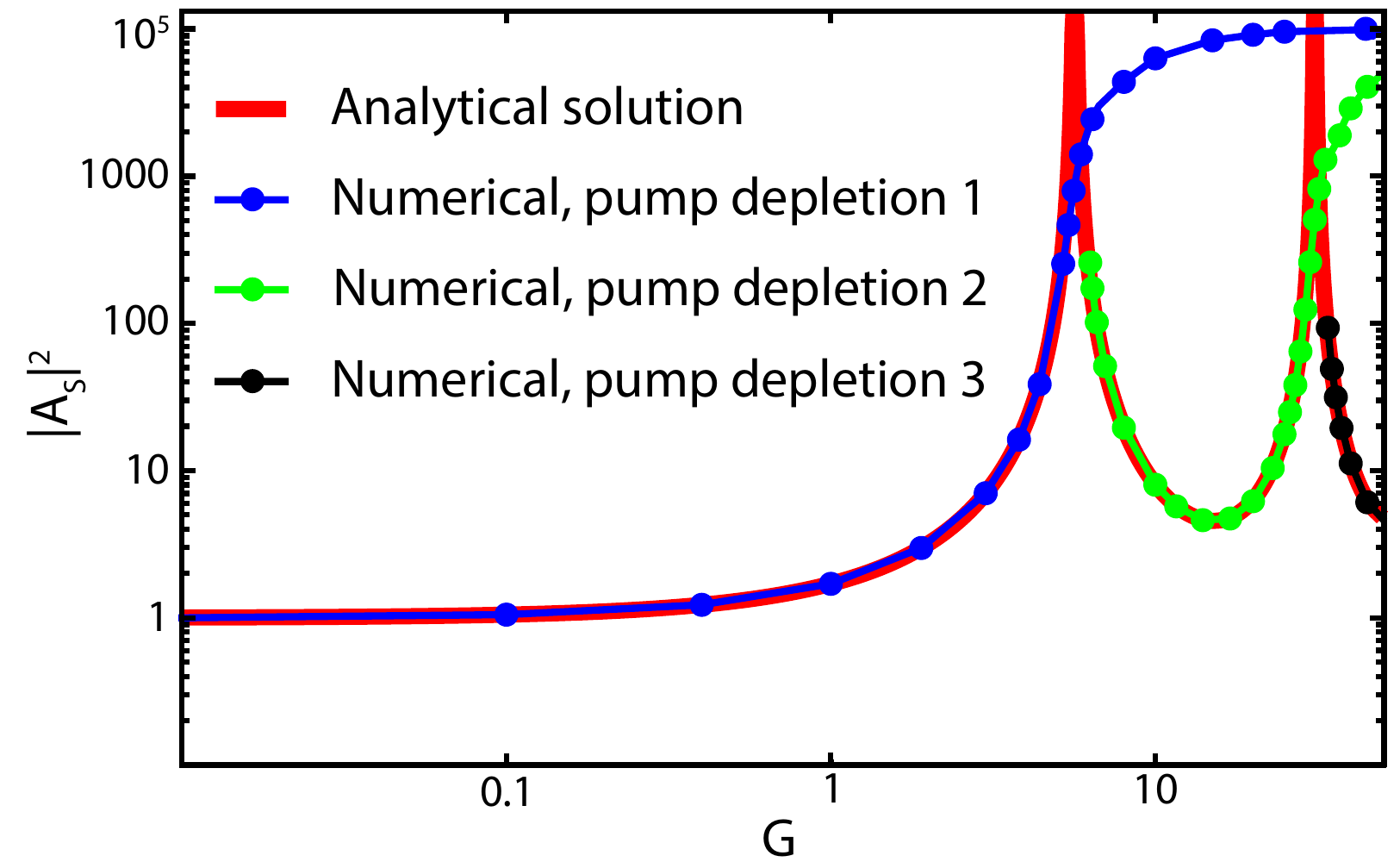}
\caption{\label{fig:amplitude} Plot of a backward SBS amplifier's Stokes amplitude squared for the analytical solution (red) and the numerical solutions (blue, green, black). For this plot, we have not included the noise Langevin terms, assumed  $\Gamma L/v_{g,b}=2\pi\times0.5$, and for the numerical solution we have included pump depletion with a pump initial condition of $\overline{A}_p(L)=100$. Due to the boundary conditions of this system, there are divergences in the analytical solution that, in the numerical solution, introduce cascading multiple solutions.}
\end{figure}

In this Appendix we examine the behavior of the backward SBS Stokes signal in the regime of relatively small phonon decay, i.e., the eigenvalues (Eq. \ref{eigen}) of our system are complex.  In Fig. \ref{fig:amplitude} we plot the backward SBS Stokes signal squared for Fig. \ref{fig:GB}a, $\Gamma L/v_{g,b}=2\pi\times0.5$. 
We see divergences appear in the analytical solution in Fig. \ref{fig:amplitude}, artifacts of the undepleted pump approximation. These divergences also appear in the noise terms, so the noise factor remains well behaved at all positive values of $\Gamma L/v_{g,b}$ and $G$. To better understand the physical behavior of the signal and the origin of these divergences, we start with this set of steady state differential equations without noise, 
\begin{align} 
  v_{g,b} \frac{\partial\overline{B}}{\partial z}&= i g_0^* \overline{A}_s^\dagger \overline{A}_p+ \frac{\Gamma}{2} \overline{B}  
  \label{Beomdeplete} 
  \\ 
   v_{g,p} \frac{\partial\overline{A}_p}{\partial z}&=   i g_0 \overline{A}_s \overline{B} 
  \label{Apeomdeplete}  
  \\
  v_{g,s} \frac{\partial\overline{A}_s}{\partial z}&=   - i g_0^* \overline{A}_p \overline{B}^\dagger. \label{Aseomdeplete} 
\end{align}
We can analytically solve this set of equations assuming pump depletion ($\overline{A}_p$ is constant) and that there is no phonon decay. The Stokes amplitude in this case is $\overline{A}_S(z)= \cos[\sqrt{G\Gamma/(4 v_{g,b} L)}z]+\sin[\sqrt{G\Gamma/(4 v_{g,b} L)}z]\tan[\sqrt{G\Gamma/(4 v_{g,b})}/L]$. Noting that $G\Gamma$ is not dependent on the phonon decay, as $G$ contains a $1/\Gamma$ dependence, we see that the origin of the divergences originate from the tangent term. The divergent behavior remains until the eigenvalues of the analytical solution $\lambda_\pm$ becomes real-valued. To examine the signal behavior including pump depletion, we use the finite difference method starting at $z=L$. The value of $\overline{A}_S(L)$ is varied until the initial condition at $z=0$ was satisfied. We can see in Fig. \ref{fig:amplitude} that the nonlinear nature of the depleted pump amplifier leads to multiple solutions to this set of equations (blue, green, and black plots). These multiple solutions cascade at each divergence that occurs in the undepleted case. Further questions on whether these additional solutions shown in Fig. \ref{fig:amplitude} are stable and can be accessed in experiment remain. As for the impact of this behavior on the noise factor, these divergences occur in the noise terms as well, meaning that that resultant SNR and noise factors are well behaved at the values of $G$ corresponding to these signal divergences. We note that the undepleted pump approximation may not be valid in the case of large Brillouin gain and significant input Stokes power. For larger values of $\Gamma L/v_{g,b}$ that result in real eigenvalues, the signal behavior is exponential, converging to $e^{G/2}.$ 

\bibliography{refs}

\end{document}